\documentclass[aps,pra,twocolumn,showpacs,superscriptaddress]{revtex4-1}
\usepackage{graphicx}  
\usepackage{dcolumn}   
\usepackage{bm}        
\usepackage{amssymb}   

\begin{document}
\title{Reentrant phases in electron-doped $\text{EuFe}_2\text{As}_2$: spin glass and superconductivity}
\author{A.~Baumgartner}
\affiliation{1. Physikalisches Institut, Universit\"at Stuttgart, Pfaffenwaldring 57, 70550 Stuttgart, Germany}
\author{D.~Neubauer}
\affiliation{1. Physikalisches Institut, Universit\"at Stuttgart, Pfaffenwaldring 57, 70550 Stuttgart, Germany}
\author{S.~Zapf}
\affiliation{1. Physikalisches Institut, Universit\"at Stuttgart, Pfaffenwaldring 57, 70550 Stuttgart, Germany}
\author{A. V.~Pronin}
\affiliation{1. Physikalisches Institut, Universit\"at Stuttgart, Pfaffenwaldring 57, 70550 Stuttgart, Germany}
\author{W. H.~Jiao}
\affiliation{Department of Physics, Zhejiang University, Hangzhou 310027, China}
\author{G. H.~Cao}
\affiliation{Department of Physics, Zhejiang University, Hangzhou 310027, China} \affiliation{Collaborative Innovation Center of Advanced Microstructures, Nanjing 210093, China}
\author{M.~Dressel}
\affiliation{1. Physikalisches Institut, Universit\"at Stuttgart, Pfaffenwaldring 57, 70550 Stuttgart, Germany}

\date{\today}

\begin{abstract}
We report evidence for a reentrant spin glass phase in electron-doped $\text{EuFe}_2\text{As}_2$ single crystals
and first traces of the superconductivity re-entrance in optics. In the close-to-optimal doped $\text{Eu}(\text{Fe}_{0.91}\text{Ir}_{0.09})_2\text{As}_2$ and $\text{Eu}(\text{Fe}_{0.93}\text{Rh}_{0.07})_2\text{As}_2$ samples two magnetic transitions  are observed below the superconducting critical temperature $T_c \approx 21$~K: the canted $A$-type antiferromagnetic order of the $\text{Eu}^{2+}$ ions sets in around $17\,\text{K}$; the spin glass behavior occurs another $2\,\text{K}$ lower in temperature. In addition, strong evidence for an additional transition is found far below the spin glass temperature.
Our extensive optical and magnetic investigations provide new insight into the interplay of local magnetism and superconductivity in these systems and elucidate the effect of the spin-glass phase on the reentrant superconducting state.
\end{abstract}

\maketitle

\section{Introduction}
Quite some time passed since in 2008 Hosono \emph{et al.} discovered
superconductivity in fluorine-doped LaFeAsO with a $T_c$ of
$26\,\text{K}$ \cite{Hosono2008} and by now a large variety of
different iron-based superconductors are found and studied
\cite{BuchIP}. Soon it was realized that in pnictides the interplay
of magnetism and superconductivity is crucial for the understanding;
the spin-density-wave phase has to be suppressed in order to develop
the superconducting phase. Eu-based iron pnictides provide an even
more interesting playground, as they allow to study the influence of
strongly magnetic Eu$^{2+}$ ions on the superconducting state \cite{SinaReview}.

The parent compound EuFe$_2$As$_2$ develops not only the
antiferromagnetic order of the itinerant iron electrons that is
typical for iron pnictides ($T_{\rm SDW} \approx 190$\,K); at low
temperatures, an additional  magnetic order of the local rare earth
moments sets in  ($T_N \approx 20$\,K) \cite{WuEu}. This local
magnetism is strong enough for a giant, indirect spin-lattice
coupling that allows a structural detwinning of these compounds by
small magnetic fields \cite{Zapf2014}. Nevertheless,
superconductivity can be induced on the same temperature scale by
chemical substitution or mechanical pressure. This leads to the
emergence of interesting phenomena such as spin-glass behavior,
reentrant superconductivity and probably spontaneous vortices.

In principle, the reentrant superconductivity is caused by a
competition between superconductivity and magnetism. Such unusual
behavior of resistivity was first predicted for superconductors
exhibiting the Kondo effect by M\"uller-Hartmann \emph{et al.}
\cite{Kondo}, but several rare earth compounds also show such a
resistivity behavior \cite{borocarbide}. In Eu-doped compounds,
external magnetic fields affect the appearance of this feature
\cite{ParamanikReentrantSC,reentrantmagneticfield,Jiao17}. Up to now it is
not fully understood why some compositions exhibit reentrant
superconductivity and others not.

Here, Eu-based 122 compounds with electron doping are investigated.
Thereby, optical, dc and ac magnetization, and dc resistivity
measurements are performed. From our study we conclude that
reentrant superconductivity is widely present in electron-doped Eu-122 compounds.
We can reveal important information on the intermediate regime, $T_{c,0} < T < T_{c,\text{on}}$, 
where the reentrant spin-glass phase affects superconductivity
until complete phase coherence is reached.

\section{Experimental Details}
The single crystals used for this study were grown by the self-flux
method \cite{Jiang09,Jiao11} and characterized by x-ray, dc
transport and susceptibility measurements. Optical investigations
are performed by normal-incidence reflection measurements in the
frequency range between $\nu=\omega/(2\pi c)=12$ and
$10\,000\,\text{cm}^{-1}$. For the far-infrared range a Fourier-transform spectrometer
Bruker IFS 113 v was utilized; the mid-infrared and near infrared data were
taken by a Bruker VERTEX 80v. In the far-infrared range
\emph{in-situ} gold evaporation technique allows us to measure at
very low frequencies. For mid- and near-infrared range an attached
Bruker Hyperion 1000 microscope provides optimal results by using a
small focus area on the sample. Temperature and field-dependent
magnetization data are obtained in three different ways by employing
a Quantum Design ac SQUID (superconducting quantum interference
device). First, the sample is cooled down without a
magnetic field applied and the data are recoreded during warming up
(zero-field cooled, ZFC). The second way is to measure upon cooling
with a field applied (field-cooled cooling curve FCC). The third
procedure records the magnetization while warming up, after the
sample was cooled down with the magnetic field present (field-cooled
heating curve, FCH). The temperature-dependent dc resistivity is
measured with the standard four-point probe technique.

In the following we focus on two electron-doped EuFe$_2$As$_2$
compounds, one with 7\%\ iron replaced by rhodium and one with 9\%\
iridium. Our magnetization measurements clearly identify the
Eu$^{2+}$ ordering at $T_N\approx 17$~K \cite{WuEu,SinaEu}; in
addition all compounds show strong evidence of the reentrant
spin-glass state\textbf{,} like it was reported for isovalent
substituted $\text{EuFe}_2(\text{As}_{1-x}\text{P}_x)_2$
\cite{SpinGlass}. This immediately tells us that the spin glass
phase is also present in electron-doped compounds and infers that
the appearance of the reentrant spin glass phase is important
for the formation of the superconducting state. Interestingly, while
$\text{Eu}(\text{Fe}_{0.93}\text{Rh}_{0.07})_2\text{As}_2$
and $\text{Eu}(\text{Fe}_{0.91}\text{Rh}_{0.09})_2\text{As}_2$
\cite{Jiao17}
exhibits
reentrant superconductivity, no signs of it are found in
$\text{Eu}(\text{Fe}_{0.91}\text{Ir}_{0.09})_2\text{As}_2$. From the
infrared results at temperatures below the onset of the
superconducting state $T_{c,\text{on}}$, we find distinct
differences in the low-frequency optical behavior for the samples
with and without reentrant superconductivity.

Besides of that, both samples show an untypical feature in the
imaginary part of the magnetic susceptibility
$\chi^{\prime\prime}_{\rm ac}$ at very low temperatures. The origin
of this feature could be either due to vortex dynamics or attributed
to a ferromagnetic transition regarding to the results
obtained in Ru-doped samples \cite{Ru}. This would be consistent with recent publications
of neutron diffraction measurements report indications of a
ferromagnetic state in $c$-direction for Co- \cite{JinCoFerro}, Ir-
\cite{AnandIrFerro, Ir12Jin} and  P-doped \cite{NandiPFerro}
samples; however they cannot rule out a small canting of the spins
which is enough to form a spin glass in the $ab$-plane. Hence it is reasonable to have a closer look at all features from the results out of the magnetization measurements.

\section{Transport Properties}
\begin{figure}[h]
    \centering
    \includegraphics[width=1.0\columnwidth]{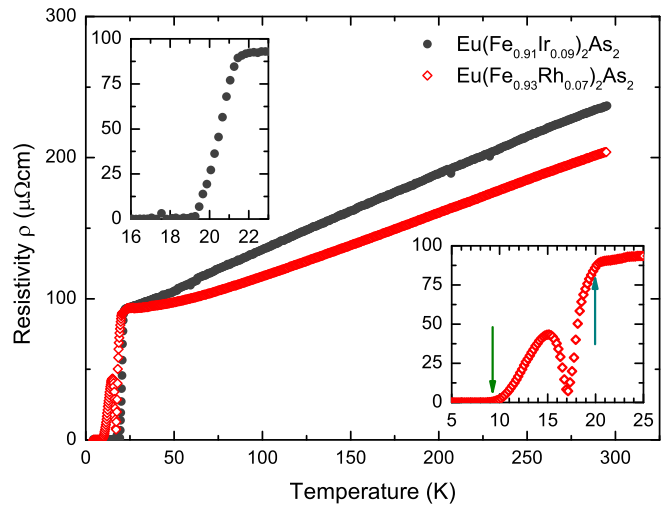}
    \caption{Temperature-dependent in-plane resistivity of $\text{Eu}(\text{Fe}_{0.91}\text{Ir}_{0.09})_2\text{As}_2$ (black circles) and $\text{Eu}(\text{Fe}_{0.93}\text{Rh}_{0.07})_2\text{As}_2$ (red diamonds); the insets highlight the superconducting transition. While for Ir-doping, $\rho_{\text{ab}}(T)$ is linear in $T$ for the normal state and shows a sharp superconducting transition, the Rh doped sample displays a quadratic temperature dependence and a resistivity re-entrance in the vicinity of the europium magnetic ordering at $T_N$.}
\label{ResAll}
\end{figure}

Figure~\ref{ResAll} displays the temperature-dependence of the
in-plane resistivity $\rho_{\rm ab}$ for both alloys,
Eu(Fe$_{0.91}$Ir$_{0.09}$)$_2$As$_2$  and
Eu(Fe$_{0.93}$Rh$_{0.07}$)$_2$As$_2$. The corresponding residual
resistivity ratios (RRR) are $\text{RRR}_{\rm Ir}=
\rho(300\,\text{K})/\rho(22\,\text{K})=2.56$ and
$\text{RRR}_{\text{Rh}}=2.24$, respectively. These values are
comparable to those reported for other Ir- and Rh-doped samples
\cite{optimaldopingIr,Ir12Jin,Jiao17} and confirm the high quality of
our crystals.

In case of
$\text{Eu}(\text{Fe}_{0.91}\text{Ir}_{0.09})_2\text{As}_2$, the
normal state resistivity depends linearly on $T$, which is typical
for unconventional superconductors close to optimal doping level
($x=0.12$) \cite{optimaldopingIr}. Such a behavior is often ascribed
to the vicinity of a quantum critical point~\cite{QCP}, but could be
also explained by spin fluctuations~\cite{Kurita}. In contrast, the
resistivity of
$\text{Eu}(\text{Fe}_{0.93}\text{Rh}_{0.07})_2\text{As}_2$ depends
quadratically on temperature between 25 and $90\,\text{K}$,
indicating Fermi liquid behavior. Therefore, this sample is probably
slightly under- or overdoped. It is interesting to recall the
behavior of BaFe$_2$As$_2$ doped with Co and Ni, where a similar
distinction was observed \cite{Barisic10}. There it was concluded
that two electronic subsystems are present with one following a
$T^2$ behavior in $\rho(T)$ up to elevated temperatures, indicating
a hidden Fermi-liquid behavior. The superconducting state evolves
out of this Fermi-liquid state.

Let us now turn to the superconducting state by looking at the
insets of Fig.~\ref{ResAll}. For the optimally doped Ir sample, a
sharp superconducting transition takes place between
$T_{c,\text{on}}\approx 21 \,\text{K}$ and $T_{c,0}\approx
18.9\,\text{K}$. In contrast,
$\text{Eu}(\text{Fe}_{0.93}\text{Rh}_{0.07})_2\text{As}_2$ exhibits
a clear resistivity re-entrance: superconductivity sets in at
$T_{c,\text{on}}\approx 19.6\,\text{K}$. However, the resistivity
does not directly drop to zero; it goes through a local minimum
around $T\approx 17\,\text{K}$, rises again to a maximum around
$T\approx 15\,\text{K}$ and reaches zero only at
$T_{c,0}=9.1\,\text{K}$. Although reentrant superconductivity for
Eu-doped compounds was reported previously, most studies considered
the effect of an external magnetic fields on this feature
\cite{ParamanikReentrantSC,reentrantmagneticfield}. It is not yet fully understood
why the reentrant behavior manifest itself in some compositions,
while not detected in others. However, when $T_c \approx T_N$ or $T_N
> T_c$, in general reentrant superconductivity behavior is
observed~\cite{Tokiwa}. This is also the reason why for samples near
optimal doping, where $T_c \gg T_N$, reentrant superconductivity is
typically not present.

\section{Optical Properties}
\subsection{Normal state}
\begin{figure}
    \centering
    \includegraphics[width=1\columnwidth]{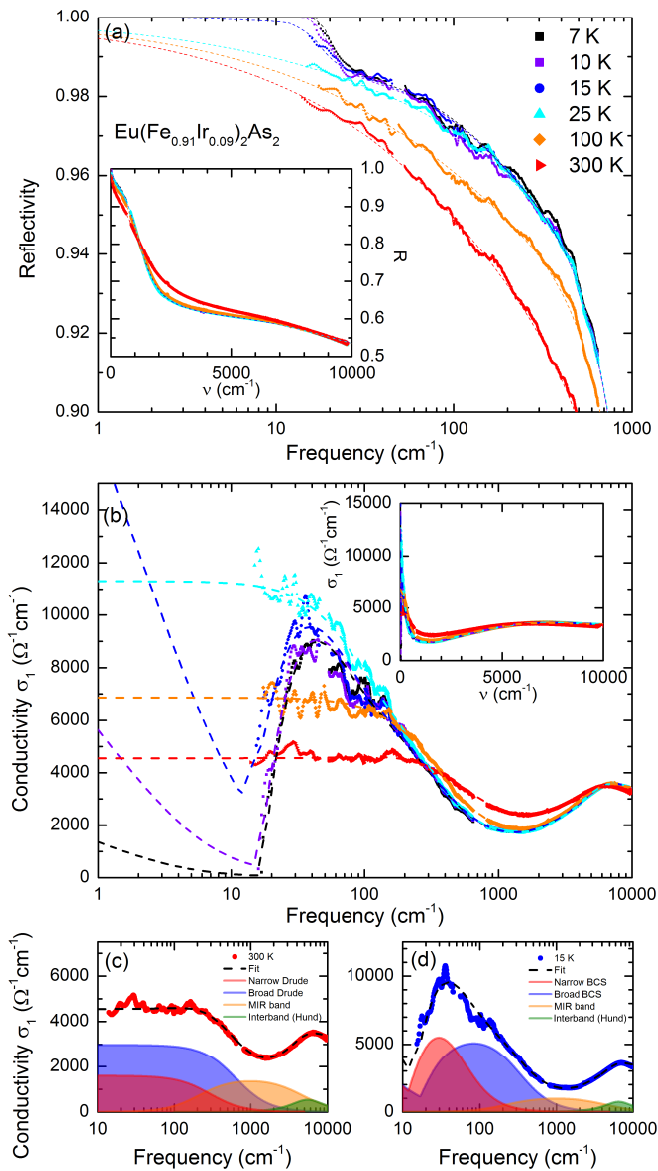}
    \caption{(a)~Optical reflectivity and (b-d)~corresponding conductivity of $\text{Eu}(\text{Fe}_{0.91}\text{Ir}_{0.09})_2\text{As}_2$ single crystals plotted for selected temperatures as indicated by different colors. The solid lines correspond to measured data, the dashed lines refer to the fits. The data are replotted on a linear scale in the insets. (a)~Below $T=25\,\text{K}$, the low-frequency reflectivity shows a pronounced upturn towards unity around $\nu=10$ to 20~cm$^{-1}$, giving evidence for the opening of the superconducting energy gaps.  (b)~Correspondingly, for $T<T_c\approx 21$~K the optical conductivity $\sigma_1(\omega)$ drops towards low frequencies. Moreover, in the normal state spectral weight is transferred from around $1000\,\text{cm}^{-1}$ to higher energies as the temperature is reduced. (c)~In the metallic state ($T=300\,{\rm K}$), the fits consist of two Drude terms (blue and red), one mid-infrared Lorentzian (orange) and one temperature-dependent Lorentzian (magenta). Note that several Lorentzians above $10\,000\,\text{cm}^{-1}$ are not shown here. (d)~Below $T_c$, the two Drude contributions are replaced by two BCS terms with two distinct energy gaps, describing the superconducting state, as exemplarily shown for $T=15\,\text{K}$. The narrow Drude exhibits the smaller gap.}
    \label{IrOptics}
\end{figure}

\subsubsection{Eu(Fe$_{0.91}$Ir$_{0.09}$)$_{2}$As$_{2}$}
In Figure~\ref{IrOptics}(a) we plot the optical reflectivity of the
optimally doped Ir sample for selected temperatures. As typical for
iron pnictides of the 122 family, no clear plasma edge can be
detected~\cite{WuEu,ElScatteringEffects}. In order to maximize the
quality of our fits, reflectivity and optical conductivity are
described simultaneously, because the complex components are linked
by the Kramers-Kronig relations. Following our previous approach
\cite{Wu10,Barisic10}, the normal state behavior is decomposed into
two Drude components (one narrow, one broad), two mid-infrared
Lorentzians, and Lorentzians located at high energies, as
illustrated in Fig.~\ref{IrOptics}(c). Since the latter ones are
located out of the measured energy range, they are considered
temperature-independent to avoid any artificial spectral weight
transfer. The spectral weight of a certain conductivity term, $\rm{SW}^{\rm{term}}$,
measures how many charge carriers $N$ contribute;
it is calculated from the optical conductivity via
\cite{DresselGruner02}
\begin{eqnarray}
\rm{SW}^{\rm{term}}=\int_0^{\infty}
\sigma^{\rm{term}}_1(\omega)\text{d}\omega=\frac{(\omega^{\rm{term}}_{\text{p}})^2}{8}=\frac{\pi
	N e^2}{2m} \quad .
\end{eqnarray}
Here $\sigma^{\rm{term}}_1(\omega)$ is the real part of the conductivity in dependence
of frequency $\omega=2\pi c \nu$; the plasma frequency of this term is indicated by
$\omega^{\rm{term}}_{\text{p}}$, $e$ is the elementary charge and
$m$ the mass of a charge carriers. Similarly, the spectral weight
can be calculated for the total optical spectrum. Here, it is
interesting consider the spectral weight as a function of the upper integration
limit -- the cut-off frequency $\omega_c$:
\begin{eqnarray}\label{eq:SW}
\rm{SW}(\omega_c)=\int_{0+}^{\omega_c}
\sigma_1(\omega)\text{d}\omega \quad ,
\end{eqnarray}
where $\sigma_1(\omega)$ is the real part of the (total) optical
conductivity, and the lower integration limit $0+$ emphasizes that
in the superconducting state we do not include the
zero-frequency $\delta$-function. While
calculating the total spectral weight hereafter, we always use
$\sigma_1(\omega)$ obtained from the fits.

\begin{figure}[b]
    \centering
    \includegraphics[width=1.0\columnwidth]{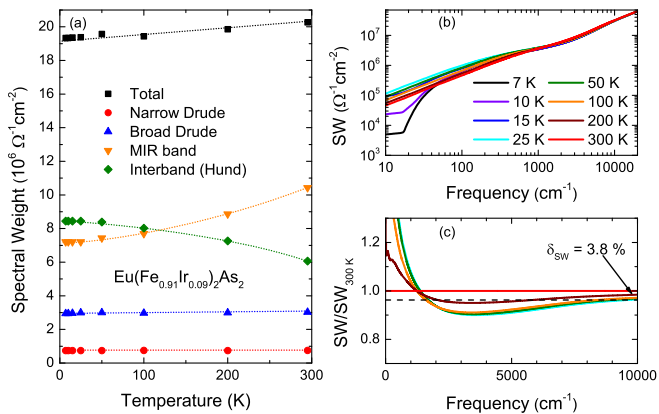}
    \caption{(a)~Temperature- and (b-c)~frequency-dependent spectral weight of $\text{Eu}(\text{Fe}_{0.91}\text{Ir}_{0.09})_2\text{As}_2$; dashed lines are guides to the eye. (a)~By fitting the spectra with Drude components with moderate scattering rates, the Drude contributions possess constant SW with temperature, and SW is transferred from the MIR band to higher energies. (b)~At $T<T_c$ , SW at low frequencies is transferred into the delta peak at $\nu=0\,\text{cm}^{-1}$. (c)~The normalized curves to $300\,\text{K}$ reveal that even at $10000\,\text{cm}^{-1}$, the spectral weight is not fully recovered with a discrepancy of $\delta_{\text{SW}}=3.8\,\%$.}
    \label{IrSpecweight}
\end{figure}
For each component of the fit, figure~\ref{IrSpecweight}(a) presents the spectral weight as a function of $T$.
The spectral weight of the narrow Drude component, often linked to the electron bands, has the smallest contribution and stays constant with temperature. This implies that no charge carriers are redistributed to other components.
The same holds for the broad Drude term, which possesses a nearly three times larger spectral weight.
This ratio is surprising, on the first glance, since the broad Drude component is commonly linked to hole bands, but the sample under investigation is electron-doped.
Therefore we should consider other possible interpretation of the two terms based on coherent and incoherent carriers \cite{Nakajima}.

Already in the conductivity data displayed in Fig.~\ref{IrOptics}(b)
it becomes  obvious that with decreasing temperature the
mid-infrared band looses spectral weight to higher energies. Indeed,
this transfer takes place up to surprisingly high energies, as can be best seen from Fig.~\ref{IrSpecweight}(b-c), which present the
spectral weight as a function of cut-off frequency $\omega_c$
[see Eq.~(\ref{eq:SW})]. As typical for iron pnictides of the
122-family, the spectral weight (down to $T=25\,\text{K}$) is not
fully recovered at $10\,000\,\text{cm}^{-1}$
\cite{ElScatteringEffects,Bing}, with a discrepancy of
$\delta=3.8\,\text{\%}$ remaining. Such a high-energy
spectral-weight transfer can be ascribed to Hund's rule
coupling~\cite{Hund}. In this case, free electrons become polarized
by localized carriers and thus do not contribute to the response of
itinerant carriers anymore, resulting in a spectral-weight transfer
of one Drude component to higher energies. As we have demonstrated
for the isovalent substituted
$\text{EuFe}_2(\text{As}_{1-x}\text{P}_x)_2$
\cite{ElScatteringEffects}, however, a fit of equal quality can also
be obtained by the present approach. Here the scattering rate of the
broad Drude term is not artificially high. The mid-infrared band
corresponds to contributions by incoherent electrons that -- as $T$
decreases -- get even more localized by Hund's rule coupling.

\subsubsection{Eu(Fe$_{0.93}$Rh$_{0.07}$)$_2$As$_2$}
\begin{figure}
    \centering
    \includegraphics[width=1\columnwidth]{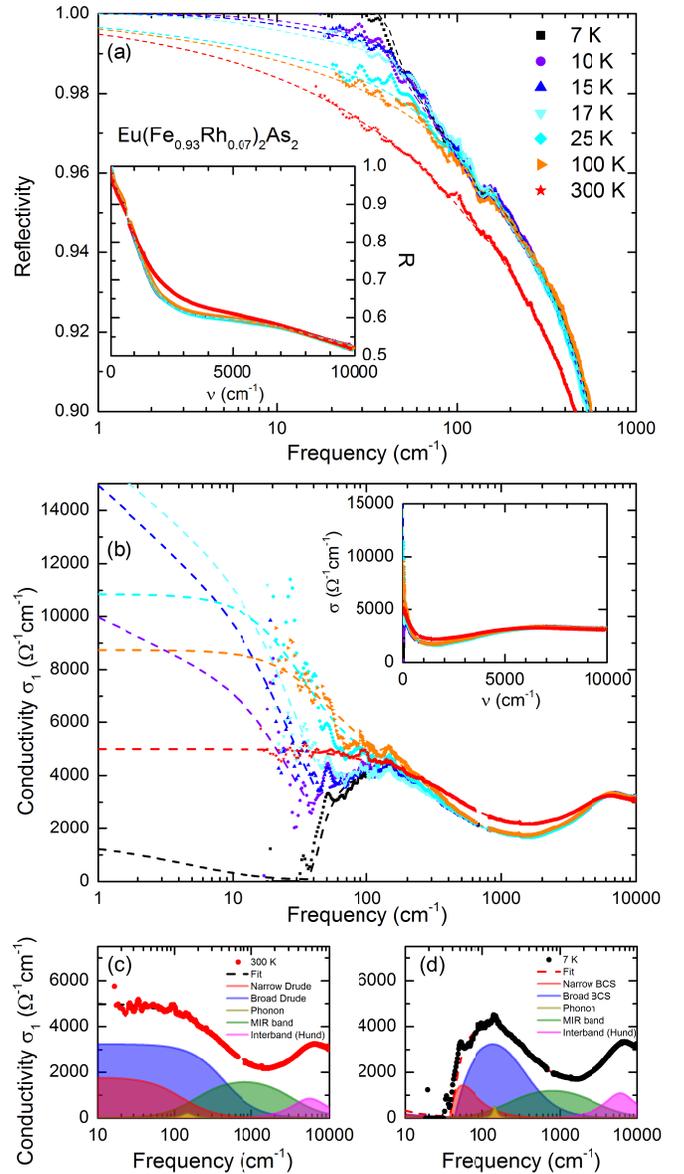}
    \caption{Temperature-dependent (a)~reflectivity and (b-d)~optical conductivity of $\text{Eu}(\text{Fe}_{0.93}\text{Rh}_{0.07})_2\text{As}_2$ at selected temperatures. The solid lines denote measurement data, the dashed lines fits and model calculations. (a)~Below $T=25\,\text{K}$, the low-frequency reflectivity shows an upturn, but only at $7\,\text{K}$ it reaches unity, as typically expected for the superconducting state.
(b)~Upon entering the superconducting phase, $\sigma_1(\omega)$ drops for frequencies below $100\,\text{cm}^{-1}$, but increases again below $60\,\text{cm}^{-1}$. Moreover, in the normal state, spectral weight is transferred from around $1000\,\text{cm}^{-1}$ to much higher energies. The inset presents the optical conductivity on a linear scale.
(c)~For the normal state ($T=300\,{\rm K}$), the fits consist of two Drude terms (blue and red), one phonon (yellow), one mid-infrared Lorentzian (green) and one temperature-dependent Lorentzian (magenta). Note that several Lorentzians above $10\,000\,\text{cm}^{-1}$ are not shown here. (d)~Below $T_c$, the two Drude contributions are replaced by two BCS terms, describing the superconducting state, as shown for $T=7\,\text{K}$, as an example. }
    \label{RhOptics}
\end{figure}

\begin{figure}
    \centering
    \includegraphics[width=1.0\columnwidth]{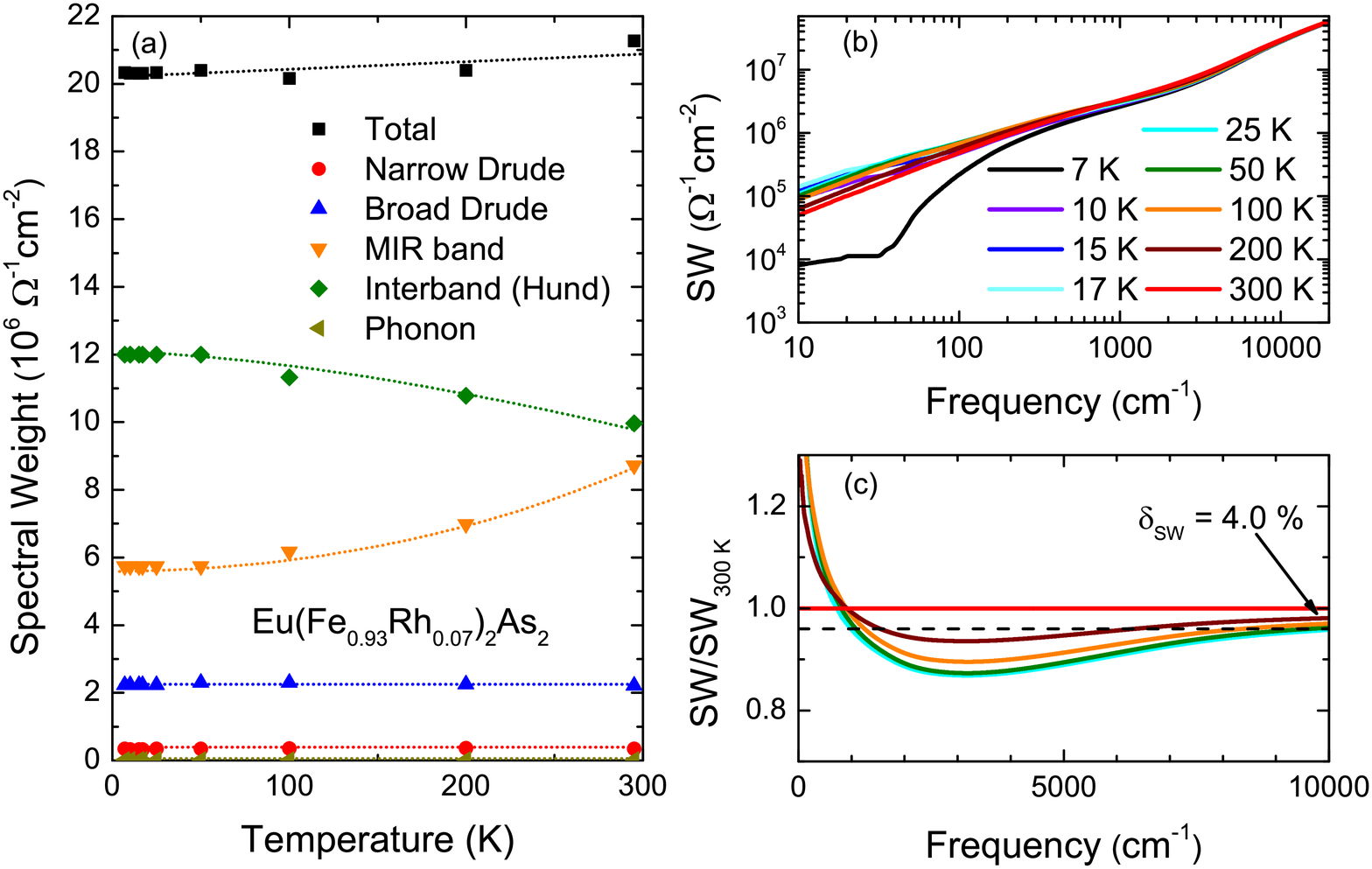}
    \caption{(a)~Temperature- and (b-c)~frequency-dependent spectral weight of $\text{Eu}(\text{Fe}_{0.93}\text{Rh}_{0.07})_2\text{As}_2$; dashed lines are guides to the eye. (a)~By fitting the spectra by Drude components with moderate scattering rates, these contributions possess constant spectral weight with temperature, and spectral weight is transferred from the mid-infrared band to higher energies. (b)~In the superconducting state at $T=7\,\text{K}$, spectral weight at low frequencies is transferred into the $\delta$-peak at zero frequency. (c)~The curves normalized to the spectral weight at $T=300\,\text{K}$ reveal that even at $10\,000\,\text{cm}^{-1}$, the spectral weight is not fully recovered with a discrepancy of $\delta_{\text{SW}}=4\,\%$.}
    \label{RhSpecWeight}
\end{figure}

Figure~\ref{RhOptics} depicts the fre\-quency- and temperature-dependent optical reflectivity and conductivity for the Rh-doped sample. In the normal state the overall behavior is very similar to the Ir-substituted crystal.
Accordingly, we use the same approach to fit the data, leading to comparable results concerning the spectral weight transfer (see Fig.~\ref{RhSpecWeight}). Again, the spectral weight is not fully recovered up to $10\,000\,\text{cm}^{-1}$, with a remaining discrepancy of $\delta=4\,\%$.

It is interesting to note the pronounced phonon at $146\,\text{cm}^{-1}$, most obvious in
Fig.~\ref{RhOptics}(c)(d).
$\text{EuFe}_2\text{As}_2$ crystallizes in the tetragonal structure of space group I4/mmm (No.139).
Torgashev calculated phonon modes at the $\Gamma$-point \cite{Torgashev} and obtained
two $E_u$ vibrations, which are infrared-active for $E\parallel x,y$.
The one at $\nu_0 = 260\,\text{cm}^{-1}$ is frequently observed in infrared
experiments \cite{WuEu,ElScatteringEffects} and also seen in related 122 iron pnictides \cite{Akrap09}.
By completely substituting Rh for Fe, this $E_u$-mode should shift by a factor of $\sqrt{m_{\text{Fe}}/m_{\text{Rh}}} \approx \sqrt{1/2}$ and is expected around $180\,{\rm cm}^{-1}$; significantly higher than the observed feature.
Thus we believe that we do observe the second $E_u$-vibration, which is predicted between $\nu=130\,\text{cm}^{-1}$ and $140\,\text{cm}^{-1}$ and has not be reported so far. The doping by Rh probably causes slight structural distortions and changes in the electron concentration leading to an enhanced intensity compared to the parent compound.

\subsection{Superconducting state}
At $T_{c,\text{on}}\approx 21$\,K and $T_{c,\text{on}}\approx
19.6$\,K, respectively, superconductivity sets in for the Ir and
Rh-doped  EuFe$_2$As$_2$ crystals, as depicted in Fig.~\ref{ResAll}.
While in the metallic phase both compounds exhibit rather similar
optical properties, their behavior in the superconducting state is
distinctively different and shall be discussed in full detail in the
following.

\subsubsection{Eu(Fe$_{0.91}$Ir$_{0.09}$)$_{2}$As$_{2}$}
\begin{figure}
    \centering
    \includegraphics[width=0.9\columnwidth]{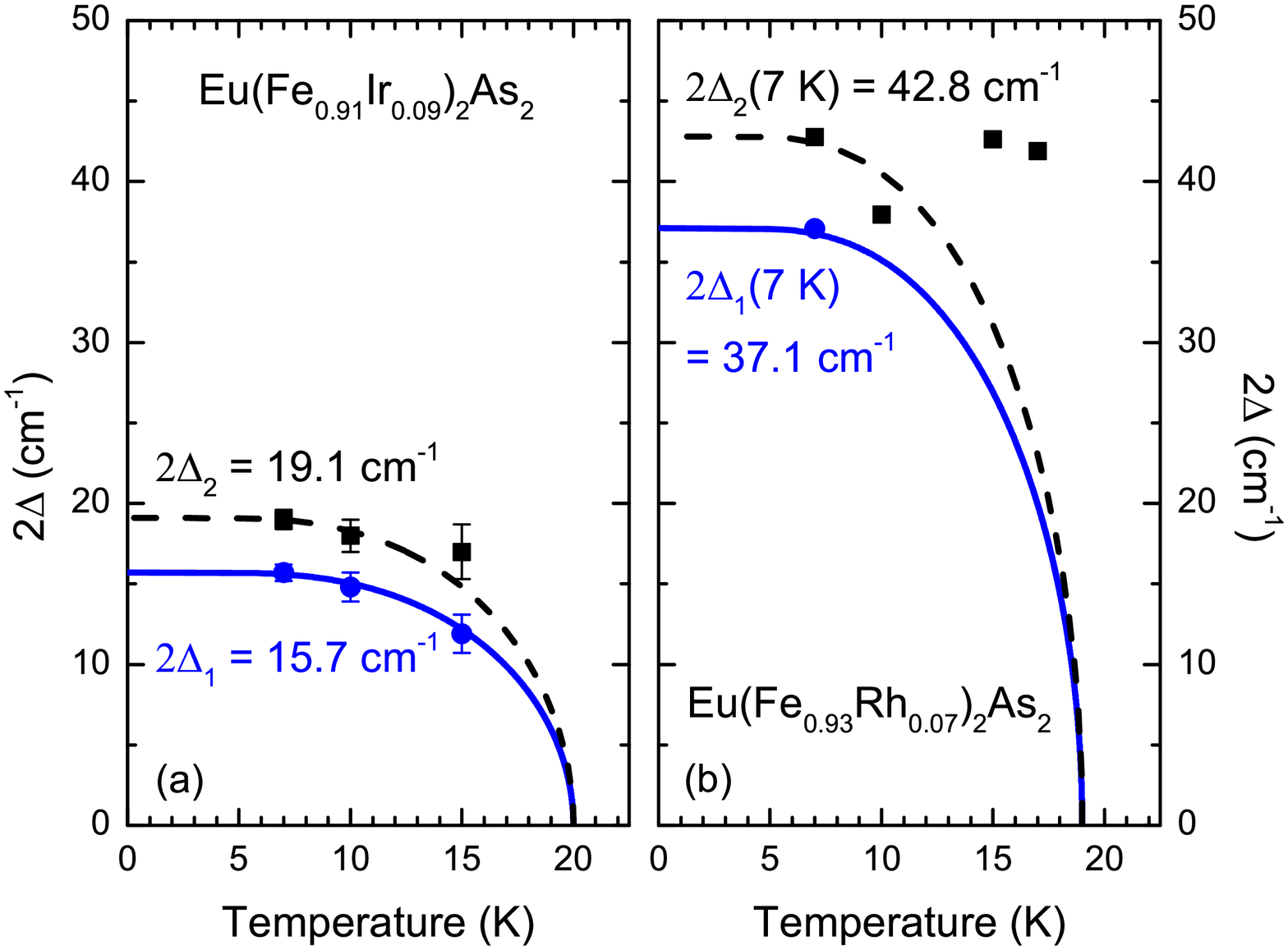}
    \caption{BCS fits of the two superconductivity gaps for (a) Eu(Fe$_{0.91}$Ir$_{0.09}$)$_2$As$_2$ and (b) Eu(Fe$_{0.93}$Rh$_{0.07}$)$_2$As$_2$. (a)~For Ir-doping the energy gaps increase with decreasing temperature below $T_c$ and can be described  well by the BCS theory. Note, however, that the ratio $2\Delta/k_BT_c$ falls far below the mean-field value of 3.5. (b)~For Rh doped samples larger gaps are observed; however, no temperature dependence can be extracted reliably.}
    \label{BCS}
\end{figure}

For the Ir-doped sample we find a clear upturn in reflectivity
around $24\,\text{cm}^{-1}$ towards unity; the common hallmark of a
superconductor with a complete energy gap at the Fermi surface
\cite{DresselGruner02,Dressel13}. In order to describe the
electrodynamic behavior below $T_c$, the two Drude terms are
replaced by two BCS contributions [Fig.\ \ref{IrOptics}(d)],
following the extended Mattis-Bardeen equations
\cite{Wu10,Pracht13}. We assume that both contributions remain
independent and simply add up. Since we are close to the limit what
can be measured by gold-evaporation technique, we restrain from applying more advanced models
\cite{Maksimov11}. As demonstrated in Fig.~\ref{IrSpecweight}(b),
the spectral weight is significantly reduced for $T < T_c$ and
transferred to the $\delta$-peak at $\nu =0\,\text{cm}^{-1}$. From
the BCS fit of the optical data, the temperature-dependent gaps are
extracted and displayed in Fig.~\ref{BCS}(a). As expected, both
energy gaps start to open simultaneously around 20~K, increase with
decreasing $T$, and  basically follow the BCS temperature
dependence. The zero-temperature extrapolation yields
$2\Delta_1=15.7\,\text{cm}^{-1}$ and
$2\Delta_2=19.1\,\text{cm}^{-1}$. These absolute values are about a
factor of 2.5 to 3 smaller than one would expect from
weak-coupling mean-field theory with a $T_c\approx
20\,\text{K}$; and also smaller than reported for other iron-based
superconductors of the 122 family \cite{Dressel11,BuchIP}. It is
interesting to compare our findings with one on P-substituted
EuFe$_2$As$_2$ compounds where no gap was recognized in the optical
response, because the samples are clean-limit superconductors
\cite{David}.

Recently, the coexistence of clean- and dirty-limit
superconductivity was proposed to account for the optical properties
of LiFeAs \cite{Dai16}. This approach might also be relevant for
Eu(Fe$_{0.91}$Ir$_{0.09}$)$_{2}$As$_{2}$, as
we do observe a clear signature of a small gap but there could be
another gap present at higher frequencies.
The gap feature is not detected in optical conductivity simply because its position is above the characteristic
scattering rates, i.e.\ the clean-limit situation is realized for
this larger gap. By applying these considerations
we could explain that the gap observed
is three times below the BCS value \cite{remark1}.

\subsubsection{Eu(Fe$_{0.93}$Rh$_{0.07}$)$_2$As$_2$}
The Rh-doped crystals do not only exhibit pronounced reentrant
superconductivity in the dc resistivity, as demonstrated in
Fig.~\ref{ResAll}, but also distinct optical properties for $T<T_c$.
In a close inspection of the frequency-dependent reflectivity,
plotted in Fig.~\ref{RhOptics}(a), clear fingerprints of the
re-entrance phase can be resolved below $40\,\text{cm}^{-1}$. At
$T=T_{c,{\rm on}}$, the onset of superconductivity, the
low-frequency reflectivity rises strongly; for instance, the changes
between $25\,\text{K}$ and $17\,\text{K}$ are much bigger than the one observed
between $100\,\text{K}$ and $25\,\text{K}$. In principle this behavior is
typical for a superconducting transition, however, the reflectivity
does not shoot up to unity immediately. Only at $T=7\,\text{K}$,
i.e. below the zero-resistance temperature $T_{c,0}$, the
reflectivity $R(\omega) = 1$ is finally reached. A similar
observation is made on the spectral-weight transfer to the
zero-frequency peak that can only be observed at the lowest
temperature.

The temperature and frequency behavior of the optical conductivity
displayed in Fig.~\ref{RhOptics}(b) basically leads to similar
conclusions. Only the curves below $T=25\,\text{K}$ exhibit a drop
similar to the Ir-analogue, but then $\sigma_1(\omega)$ increases
again for decreasing frequencies due to a strong in-gap absorption.
Since there are no explicit models for the frequency-dependence of a
re-entrance superconductor, we follow the procedure described above
and fit the data by replacing the Drude with a BCS term as the
superconducting state is entered. While the larger gap around
$42\,\text{cm}^{-1}$ is present already below $T\approx 18$~K, it
does not change much as $T$ is lowered. Only at $T=7\,\text{K}$ the
narrow term exhibits an energy gap, with $37.1\,\text{cm}^{-1}$; see
Fig. \ref{BCS}(b).

Similar temperature-independent optical gaps --~i.e.\
gap-related features seen in optical experiments~-- were
previously reported in $\text{Ba}(\text{Fe}_{0.92}\text{Co}_{0.08})_2\text{As}_2$ single
crystals \cite{LoboGap} and $\text{LaFeAsO}_{1-x}\text{F}_x$ thin
films \cite{XiGap}.

Up to now, the origin of such behavior in
pnictides is not fully clarified, though for the
$\text{LaFeAsO}_{1-x}\text{F}_x$ films it was explained as being due
to an interplay of the BCS-gap optical feature and an absorption
mode situated at frequencies just slightly above the
gap~\cite{XiGap}. A similar, but even more counterintuitive
$T$ dependence of the apparent optical gap was reported in
$\text{MgB}_2$~\cite{PimenovGap}: there the minimum in
$\sigma_{1}(\omega)$ decreased as $T \rightarrow 0$. This
behavior was explained by the multiband nature of
$\text{MgB}_2$: the smaller gap dominates the optical response at
the lowest temperature and the higher-frequency feature due to the larger
gap is not well pronounced in the conductivity spectra. As $T$
increases, the transition across the smaller gap becomes thermally
saturated and the conductivity minimum gradually shifts to higher
frequencies, revealing the optical feature related to the larger
gap~\cite{PimenovGap}. It seems plausible that multiband effects
also affect the temperature dependence of the optical gaps in
pnictides.

The first observation of a superconducting energy gap in the reentrant phase enables us to go one step further,
as it provides important information on the intermediate phase.
The condensate forms right below $T_{c,\text{on}}$, although the density of
superconducting electrons is very low. Since there is no superconducting path formed through the sample,
the dc resistivity is not zero. Only below $T_{c,0}=9.1\,{\rm K}$, when phase coherence is fully reached,
$\rho(T)$ vanishes and a gap opens in the narrow conductivity term.

Our observations might be explained by scattering processes. Defects created by irradiation are known to quickly close the small superconducting gap with only minor effect on $T_c$ \cite{Schilling16}. Contrary to that
Li \emph{et al.}~\cite{splusminus} suggested that in case of $\text{s}_{\pm}$ symmetry, magnetic impurities act as interband scatterers, which hardly affect the transition temperature and do not break Cooper pairs.
However, intraband magnetic scattering leads to pair breaking \cite{splusminus}.
Accordingly, superconductivity is not destroyed as long as $1/\tau_{\text{inter}}>1/\tau_{\text{intra}}$.
We suggest that the different $\text{Eu}^{2+}$ transitions lead to dominant intraband scattering at high $T$, but favor interband scattering in the superconducting phase at low temperatures.
When $T_c\approx T_N$, Cooper pairs begin to condense, as evident in optics,
but intraband magnetic scattering destroys the phase coherence
causing  a finite resistivity. When the spin-glass transition sets in, interband magnetic scattering prevails,
permitting phase coherence and a fully developed superconducting state.
Obviously the resistivity does not vanish immediately at the spin-glass transition;
this can be simply explained by the inherent nature of a glass-like transition, which develops smoothly over a certain temperature range and does not appear abruptly.

While for the intermediate regime the presented scenario seems to be the most likely explanation of our findings,
we recall that an increase in the low-frequency conductivity $\sigma_1(\omega)$ was previously taken
as indication of pseudogap formation in iron-based superconductors~\cite{Moon12}.
Moreover, in cuprates the crossover between pseudogap and superconducting gap is known to take place at similar  energy scale. Since we are very close to the limit of our resolvable measurement range, we cannot to rule out that our observations are caused by a pseudogap-like formation, where the complex interaction of different magnetic transitions have an impact on the condensate of the Cooper pairs.



\section{Magnetic properties}
In the parent compound $\text{EuFe}_2\text{As}_2$, the $\text{Eu}^{2+}$ magnetic moments order below $T_N=19\,\text{K}$ in an \emph{A}-type antiferromagnetic structure; i.e.\ while the spins are aligned ferromagnetically within the $ab$-plane, neighboring layers are coupled antiferromagnetically~\cite{ShuaiAFM}. For $\text{EuFe}(\text{As}_{1-x}\text{P}_x)_2$, the isovalent substitution of As with P leads to a canting of the spins out of the $ab$-plane, and therefore to a ferromagnetic net component along the $c$-direction~\cite{SinaEu}. As the canting increases with further P substitution, also the competition between ferromagnetism and antiferromagnetic  Ruderman-Kittel-Kasuya-Yoshida (RKKY) exchange increases. Therefore, an additional magnetic phase --~the reentrant spin glass phase~-- sets in at a certain amount of phosphorous with $T_{\text{SG}} < T_N$~\cite{SpinGlass}. As this phase might be the key in understanding how superconductivity coexists with the local magnetism in Eu-based iron pnictides, we carried out extensive magnetic investigations on our electron-doped EuFe$_2$As$_2$ compounds, Eu(Fe$_{0.91}$Ir$_{0.09}$)$_{2}$As$_{2}$ and
Eu(Fe$_{0.93}$Rh$_{0.07}$)$_2$As$_2$.

\subsection{Evidence for a spin glass}
The most prominent feature identifying a spin-glass phase is the time dependence of the dc magnetization,
together with a distinct frequency dependence of the ac-susceptibility known as Vogel-Fulcher behavior~\cite{Binder}.
However, in superconductors also vortex dynamics leads to a time-dependent magnetization, hence complicating the analysis~\cite{Zhou}.
Nevertheless, as demonstrated  previously for $\text{EuFe}(\text{As}_{1-x}\text{P}_x)_2$  \cite{SpinGlass},
both phases can be separated by their opposite signs in the time-dependent behavior
(for instance in the FCH-FCC curve, the difference between field-cooled heating and field-cooled cooling curves).
Following the procedure developed for the isovalent substitution,
in the following we will provide evidence that these phases are present in electron-doped samples, too.

\subsubsection{$\text{Eu}(\text{Fe}_{0.91}\text{Ir}_{0.09})_2\text{As}_2$}
\begin{figure}
    \centering
    \includegraphics[width=1.0\columnwidth]{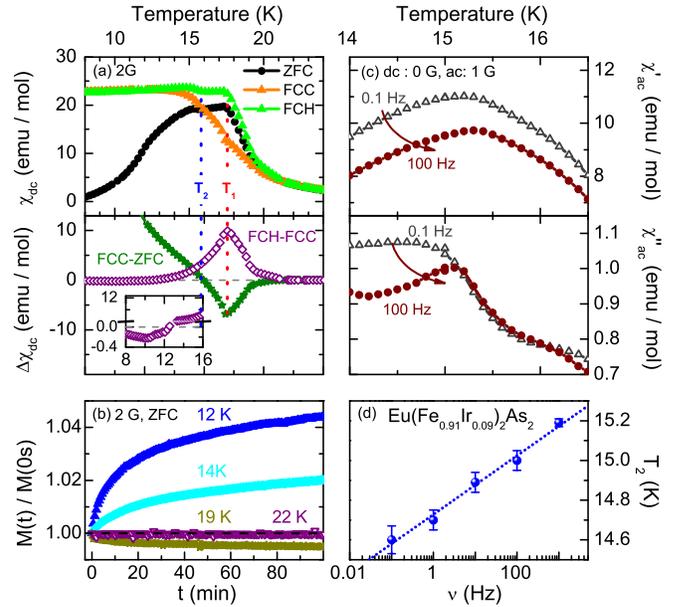}
    \caption{(a-b) $ab$-plane dc and (c-d) ac-susceptibility of Eu(Fe$_{0.91}$Ir$_{0.09}$)$_2$As$_2$ as a function of temperature. (a)~While the upper frame shows the ZFC, FCC, and the FCH run, the lower one highlights the difference of the cooling and heating cycles. Thermal hysteresis is evidenced by a finite FCC-ZFC, while a non-zero value of FCH-FCC denotes time-dependent behavior. The temperatures $T_1$ and $T_2$ marked by the dotted lines correspond to two magnetic transitions. (b)~Time-dependent measurements reveal two processes with opposite time-dependence. Below $T_c$, the magnetization decreases with time; at low temperatures, it increases. (c)~Below $T_2$, ac magnetic measurements show a frequency dependence in $\chi^{\prime}_{\text{ac}}(T)$ and  $\chi^{\prime\prime}_{\text{ac}}(T)$. (d)~The peak position in  $\chi^{\prime\prime}_{\text{ac}}(T)$ can be well described by a Vogel-Fulcher fit.}
    \label{IrPlotMag1}
\end{figure}
For the Ir-doped sample, the characteristic magnetization data are presented in Fig.~\ref{IrPlotMag1}.
The dc susceptibility [panel (a)] looks quite similar to the one obtained in Ref.~\onlinecite{SpinGlass}: the double hump-like feature in the ZFC and FCH curve already indicates both magnetic transitions:
the Eu-ordering corresponding to the first hump at $T_1=T_N=17.5\,\text{K}$, and the spin-glass phase indicated by the second hump at $T_2=T_{\text{SG}}=15.9\,\text{K}$, with $T_2<T_1$.
The measurements are performed at very low magnetic field ($\mu_0 H =2\,\text{G}$) since the spin-glass phase is already suppressed by $500\,\text{G}$.

Both, a thermal hysteresis (evidenced by the finite difference between field-cooled heating and zero-field cooled cooling curves: FCC-ZFC), and time-dependent processes (visible in FCH-FCC $\neq 0$) set in at the superconducting transition, which can be ascribed to vortex dynamics. At $T_1$, a peak occurs in the FCH-FCC curve, and a dip in FCC-ZFC, meaning that the onset of Eu ordering influences the superconducting phase. At lower $T$, both differential curves show a sign change, which is typical for the spin-glass phase~\cite{SpinGlass}. For the FCH-FCC curve, the sign change is not directly at $T_2$, but at slightly lower temperatures, probably due to the dominant superconducting state which masks the negative contribution.
Time dependent measurements in the ZFC run further prove two distinct contributions to the overall time dependence as plotted in Fig.~\ref{IrPlotMag1}(b). It is important to note that any time dependence disappears at $T>T_c$, ruling out possible measurement artifact. For $T<T_c$ the normalized magnetization decreases with time, while for temperatures $T<T_2$, $M(t)$ increases significantly; we ascribe the stronger positive contribution to the spin-glass phase.
Finally, the typical spin glass behavior can also be identified in ac susceptibility measurements with a very small amplitude of $1\,\text{G}$ and without dc offset: here, a peak appears directly below $T_2$ in $\chi_{\text{ac}}^{\prime}(T)$ and $\chi_{\text{ac}}^{\prime\prime}(T)$ [cf. Fig.~\ref{IrPlotMag1}(c)]. With increasing frequency, the peak shifts to higher temperatures. The peak position in $\chi_{\text{ac}}^{\prime\prime}$ follows a Vogel-Fulcher behavior, verifying the identification of this transition to the spin glass phase.
\begin{figure}
    \centering
    \includegraphics[width=0.74\columnwidth]{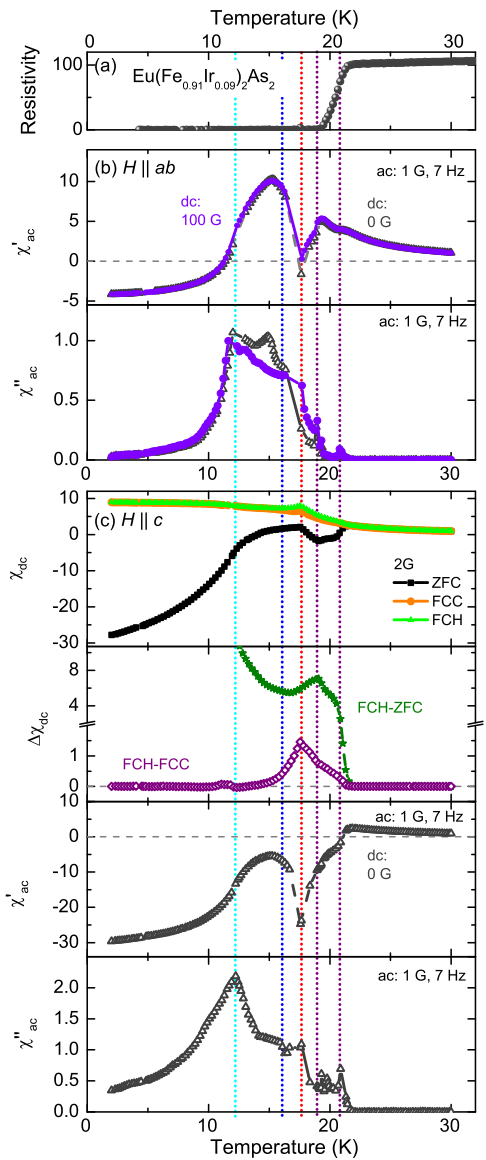}
    \caption{(a)~Resistivity and (b-c)~magnetization of $\text{Eu}(\text{Fe}_{0.91}\text{Ir}_{0.09})_2\text{As}_2$ as a function of temperature (in emu/mol); the dotted lines are guides to the eye and mark the different transitions: in additional to the temperatures $T_1$ (red) and $T_2$ (blue), we also marked the response of the superconducting transition (purple) and another feature (cyan) not assigned yet. (a)~The entrance to the superconducting state is clearly visible in the resistivity. (b)~In-plane (\emph{ab}) susceptibility data $\chi_{\text{ac}}^{\prime}(T)$ (top) and $\chi_{\text{ac}}^{\prime\prime}(T)$ (bottom) for a dc offset of $\mu_0 H=0\,\text{G}$ (grey) and $100\,\text{G}$ (purple). $\chi^{\prime\prime}_{\text{ac}}(T)$ shows a wide variety of features: with decreasing temperature, a non-zero contribution develops below $T_{c,\text{on}}$ with a peak at approximately $21\,\text{K}$; a second peak denotes $T_{c,0}=18.9\,\text{K}$ (purple). At $T_1=17.5\,\text{K}$ a minimum in $\chi^{\prime}_{\text{ac}}(T)$ and a slope change in $\chi^{\prime\prime}_{\text{ac}}$ represents the antiferromagnetic transition ($T_1$, red). Below $T_2=15.9\,\text{K}$, a further peak develops in both contributions due to the spin-glass transition ($T_2$, blue). (c)~Measurements with $H \parallel c$ basically reveal the same transitions, however, there is no clear sign of the spin-glass transition. At $T=12\,\text{K}$ a distinct peak in $\chi^{\prime\prime}_{\text{ac}}(T)$ and a slope change in $\chi^{\prime}_{\text{ac}}(T)$ marks a possible unknown phase (cyan).}
    \label{IrPlotMag2test}
\end{figure}

In the following we will discuss in more detail the imaginary part of the ac susceptibility, because it contains several different contributions, as best seen in a comparison between ac-, dc susceptibility and resistivity data [Fig.~\ref{IrPlotMag2test}(a) and (b)]. The onset of superconductivity below $T<T_{c,\text{on}}\approx 21\,\text{K}$ leads to non-zero values in $\chi_{\text{ac}}^{\prime\prime}(T)$, corresponding to the ZFC-FCH splitting due to screening currents in the dc magnetization [Fig.~\ref{IrPlotMag1}(a)]. A narrow sharp peak in $\chi_{\text{ac}}^{\prime\prime}(T)$ and a small dip in $\chi_{\text{ac}}^{\prime}(T)$ at $T_{c,0}$ are caused by a maximum in the critical current density, known as peak effect~\cite{PeakEffect}. Around $T_1=17.5\,\text{K}$, a minimum in $\chi_{\text{ac}}^{\prime}(T)$ coincides with the first peak in the ZFC-curve, marking the Eu$^{2+}$ antiferromagnetic transition; it results from a destruction of the superconducting state by the local Eu-magnetism. Below $T_2=15.9\,\text{K}$, a peak in $\chi_{\text{ac}}^{\prime}(T)$ and $\chi_{\text{ac}}^{\prime\prime}(T)$ can be ascribed to the spin-glass phase; it is almost suppressed by a dc magnetic offset of only $100\,\text{G}$.

Additional information can be extracted from Fig.~\ref{IrPlotMag2test}(c) where the magnetization results are displayed for applied fields along the $c$-direction. The most prominent feature for this orientation is the strong negative values for the ZFC magnetization reached at low temperatures; this is well known to result from the anisotropy of iron pnictides~\cite{SFdensity}. In general, screening currents form in the superconducting plane when a magnetic field is applied along the perpendicular direction. In iron pnictides, the $ab$-plane shows much higher superfluid densities than along the $c$-direction. Therefore, screening currents can develop much more efficiently within the $ab$-plane, visible as a negative magnetization for $H \parallel c$. Otherwise, for $H \parallel c$
superconductivity manifests in a way very similar to the in-plane direction: a thermal hysteresis at $T<T_{c,\text{on}}$, a peak at $\chi^{\prime\prime}_{\text{ac}}(T_{c,0})$, and the peak effect is visible.

The spin-glass phase, in contrast, is barely visible along the $c$-direction; for instance, in the ZFC, FCC, and FCH curves, only one hump can be detected, which corresponds to the antiferromagnetic transition at $T_1$. A small hump is visible in the ac measurements, however, without any resolvable time- or frequency-dependence; therefore we ascribe it to a measurement artifact due to non-perfect alignment of the sample within the SQUID.
For temperatures $T<T_1$, the FCC and FCH curves are rather flat; this is typical for an $A$-type antiferromagnet when the out-of-plane direction is probed. A small increase of these curves at lower temperatures, as well as a weak peak at $\chi^{\prime\prime}_{\text{ac}}(T_1)$ probably results from canting of the Eu$^{2+}$ moments along the $c$-direction; similar observations have been reported for P-substituted samples~\cite{SinaEu}. Unfortunately, in the superconducting specimen its contribution is too weak for a clear identification.

\subsubsection{$\text{Eu}(\text{Fe}_{0.93}\text{Rh}_{0.07})_2\text{As}_2$}
\begin{figure}
    \centering
    \includegraphics[width=1.0\columnwidth]{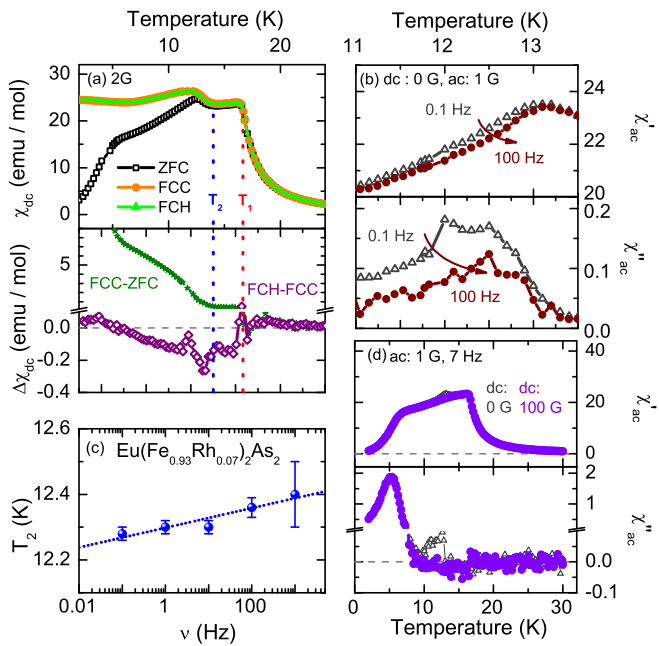}
    \caption{(a)~$ab$-plane static susceptibility and (b-d)~ac susceptibility of $\text{Eu}(\text{Fe}_{0.93}\text{Rh}_{0.07})_2\text{As}_2$ plotted as a function of temperature (in emu/mol). (a)~While the upper figure shows the ZFC, FCC and the FCH run, the lower one highlights the difference of the cooling and heating cycles. Thermal hysteresis is evidenced by a finite FCC-ZFC, while a non-zero value of FCH-FCC denotes time-dependent behavior. The temperatures $T_1$ and $T_2$ marked by the dotted lines correspond to two magnetic transitions. (b)~Below $T_2$, ac magnetic measurements show a frequency dependence in $\chi^{\prime}_{\text{ac}}(T)$ and  $\chi^{\prime\prime}_{\text{ac}}(T)$. (c)~The peak position in $\chi^{\prime\prime}_{\text{ac}}(T)$ can be well described by a Vogel-Fulcher fit. (d)~In-plane ($ab$) susceptibility data $\chi^{\prime}_{\text{ac}}(T)$ (top) and $\chi^{\prime\prime}_{\text{ac}}(T)$ (bottom) for a dc offset of $\mu_0 H= 0\,\text{G}$ (grey) and $100\,\text{G}$ (purple). For an offset field of $100\,\text{G}$ the spin glass transition seems already fully suppressed.}
    \label{RhMag}
\end{figure}

Figure~\ref{RhMag} presents the results of the magnetization measurements for the Rh-doped sample. In this case,
the sample mass was at the sensitivity limit of the SQUID leading to much noisier curves than for the Ir specimen. However, the main features are very similar, as we will discuss in the following. The onset of superconductivity is barely visible but leads to slightly non-zero, positive FCH-FCC values. A hump in $\chi_{\text{dc}}(T)$ at $T_1=16.6\,\text{K}$ marks the antiferromagnetic order; it corresponds to a weak peak in the FCH-FCC curve. At lower $T$ the double hump-like feature in the ZFC, FCC and FCH curves, as well as negative values of FCH-FCC indicate the presence of the reentrant spin glass phase. In this case, as the superconducting state is weaker, the minimum of FCH-FCC appears much closer to $T_2$ than in the Ir sample. Again, the frequency-dependent ac susceptibility shows for increasing ac fields a shifting of the peaks in $\chi^{\prime}_{\text{ac}}(T)$ and $\chi^{\prime\prime}_{\text{ac}}(T)$ to higher $T$, shown in Fig.~\ref{RhMag}(b); the peak position (fitted with a Gaussian distribution) also follows Vogel-Fulcher behavior as demonstrated in Fig.~\ref{RhMag} (c).

The poor signal-to-noise ratio does not allow us to extract here as much information from $\chi^{\prime\prime}_{\text{ac}}(T)$ as for $\text{Eu}(\text{Fe}_{0.91}\text{Ir}_{0.09})_2\text{As}_2$;
nevertheless, for the sake of completeness, we present the results in Fig.~\ref{RhMag}(d) for a large temperature range. In contrast to the Ir-doped sample, for $\text{Eu}(\text{Fe}_{0.93}\text{Rh}_{0.07})_2\text{As}_2$  $\chi^{\prime}_{\text{ac}}(T)$ does not reach negative values, indicating a stronger $\text{Eu}^{2+}$ magnetism and/or a weaker superconductivity phase -- which would be both consistent with speculations that this sample is actually underdoped. For an offset of $\mu_0 H=0\,\text{G}$, a peak is located below $T_2$ in both, $\chi^{\prime}_{\text{ac}}(T)$ and $\chi^{\prime\prime}_{\text{ac}}(T)$, corresponding to the spin-glass phase. For a constant magnetic-field offset of $\mu_0 H=100\,\text{G}$, it is already completely suppressed. Similar to the Ir sample, an additional pronounced peak appears in $\chi^{\prime\prime}_{\text{ac}}(T)$ at lower temperatures.

\subsection{Indications for an additional magnetic transition}
From our ac susceptibility measurements over a broad temperature range, we could identify a very pronounced peak in $\chi_{\text{ac}}^{\prime\prime}(T)$, which also appears as an inflection point in $\chi_{\text{ac}}^{\prime}(T)$ and in the ZFC curve. This feature can be seen even better when the magnetic field is applied along the $c$-direction. While a dc-offset of $\mu_0 H=100\,\text{G}$ completely suppresses the spin-glass peak in $\chi_{\text{ac}}^{\prime\prime}(T)$, this additional feature is barely affected by this offset.
For $\text{Eu}(\text{Fe}_{0.91}\text{Ir}_{0.09})_2\text{As}_2$ the feature appears around $12\,\text{K}$; for the Rh-doped sample, at $5\,\text{K}$. There are two possibilities to explain the observation of this signal. A ferromagnetic canting or phase would lead to a signal in $\chi^{\prime\prime}_{\text{ac}}(T)$.
In Ref.~\onlinecite{SinaReview} an extensive discussion is presented why the canting of the moments does not take place directly at $T_N$, but develops with decreasing temperature. This would explain the observation of an additional phase in Ru-doped compounds~\cite{Ru}. Alternatively, the feature could originate due to complex vortex dynamics.
In order to resolve this issue, further investigations on overdoped samples should be carried out, which do not show any traces of superconductivity.

\section{Conclusion}
Transport and magnetic investigations on electron-doped
$\text{Eu}(\text{Fe}_{0.91}\text{Ir}_{0.09})_2\text{As}_2$ and
$\text{Eu}(\text{Fe}_{0.93}\text{Rh}_{0.07})_2\text{As}_2$ single
crystals revealed the presence of the reentrant spin-glass phase.
The properties resemble the behavior observed in isovalent
substituted $\text{EuFe}_2(\text{As}_{1-x}\text{P}_x)_2$ compounds
\cite{SpinGlass}. Since the two materials contain different dopants,
we suggest that the reentrant spin-glass phase is present in all
electron-doped compounds. For the Ir-doped sample, a very strong
feature in the imaginary part of the ac susceptibility was
found, which might be explained by vortex dynamics; however, future
investigations on different doping levels are necessary to clarify
its origin.

Optical measurements revealed that at low-frequencies
the normal-state optical response of each compound is dominated by two
Drude-like components with different scattering rates. At the lowest
temperatures, in the superconducting state, the Drude components
become gapped and the optical spectra can be best
described by two full-gap dirty-limit BCS terms, with the gap values
being close to each other for both the terms in each system.
In $\text{Eu}(\text{Fe}_{0.93}\text{Rh}_{0.07})_2\text{As}_2$, the
traces of reentrant superconductivity are seen in the optical
spectra: in the reentrant phase only one
basically temperature-independent gap can be resolved,
while a second superconducting gap gets merely visible at the lowest temperatures
when zero-resistivity is finally reached.
We explain this behavior as the effect of magnetism: the superconducting state can fully develop only when the spin-glass phase is formed.

\section{Acknowledgment}
We thank U. S. Pracht for fruitful discussions and G. Untereiner for expert experimental support. This work was supported by the Deutsche Forschungsgemeinschaft (DFG SPP 1458, DR228/42-1 and DR228/44-1).


\begin{thebibliography}{99}
\bibitem{remark1}
We should note that it might not be appropriate to identify
the narrow and broad Drude components with different bands.
Also the clean and dirty limit superconductivity
in Eu(Fe$_{0.91}$Ir$_{0.09}$)$_{2}$As$_{2}$,
Eu(Fe$_{0.93}$Rh$_{0.07}$)$_2$As$_2$ and possibly in other pnictides
should not simply be assigned to distinct bands
as intra- and interband scattering processes might be of importance.
\end{thebibliography}
\end{document}